\documentclass[journal=jctcce,manuscript=article]{achemso}
\setkeys{acs}{usetitle=true}

\usepackage[version=3]{mhchem} 

\usepackage{amsmath,amssymb}
\usepackage{graphicx,color,subfigure}
\usepackage{amsfonts}
\usepackage[outdir=./]{epstopdf}
\usepackage[linesnumbered,ruled]{algorithm2e}
\usepackage{comment}
\usepackage{placeins}
\usepackage{fixltx2e}

\allowdisplaybreaks

\usepackage[version=3]{mhchem} 

\newcommand{\Or}{\mathcal{O}}
\newcommand{\C}{\mathcal{C}}

\newcommand{\norm}[1]{\left\lVert#1\right\rVert}

\newcommand{\ud}{\,\mathrm{d}}

\newcommand{\wt}[1]{\widetilde{#1}}

\newcommand{\hr}[1]{\hat{\vr}_{#1}}

\newcommand{\ph}{\phantom{ }}
\newcommand{\R}{\mathbb{R}}

\newcommand{\bvec}[1]{\mathbf{#1}}
\newcommand{\vc}{\bvec{c}}
\newcommand{\vr}{\bvec{r}}

\definecolor{MatlabY}{rgb}{0.929,0.694,0.125}

\author{Kun Dong}
\email{kd383@cornell.edu} \affiliation{Center for Applied
Mathematics, Cornell University, Ithaca, New York 14853, United
States}

\author{Wei Hu}
\email{whu@lbl.gov} \affiliation{Computational Research Division,
Lawrence Berkeley National Laboratory, Berkeley, California 94720,
United States}

\author{Lin Lin}
\email{linlin@math.berkeley.edu} \affiliation{Department of
Mathematics, University of California, Berkeley, California 94720,
United States} \alsoaffiliation{Computational Research Division,
Lawrence Berkeley National Laboratory, Berkeley, California 94720,
United States}

\title[Density fitting with centroidal Voronoi tessellation]
{Interpolative Separable Density Fitting through Centroidal Voronoi
Tessellation With Applications to Hybrid Functional
Electronic Structure Calculations}

\begin{document}

\begin{abstract}
The recently developed interpolative separable density fitting
(ISDF) decomposition is a powerful way for compressing the redundant
information in the set of orbital pairs, and has been used to
accelerate quantum chemistry calculations in a number of contexts.
The key ingredient of the ISDF decomposition is to select a set of
non-uniform grid points, so that the values of the orbital pairs
evaluated at such grid points can be used to accurately interpolate
those evaluated at all grid points.  The set of non-uniform grid
points, called the interpolation points, can be automatically
selected by a QR factorization with column pivoting (QRCP)
procedure. This is the computationally most expensive step in the
construction of the ISDF decomposition. In this work, we propose a
new approach to find the interpolation points based on the
centroidal Voronoi tessellation (CVT) method, which offers a much
less expensive alternative to the QRCP procedure when ISDF is used in
the context of hybrid functional electronic structure calculations. The
CVT method only uses information from the electron density, and can be
efficiently implemented using a K-Means algorithm. We find that this new method achieves
comparable accuracy to the ISDF-QRCP method, at a cost that is
negligible in the overall hybrid functional calculations. 
For instance, for a system containing $1000$ silicon atoms simulated
using the HSE06 hybrid functional on $2000$ computational cores, the cost of QRCP-based method for finding
the interpolation points costs $434.2$ seconds, while the CVT procedure
only takes $3.2$ seconds.  We also find that the ISDF-CVT method also
enhances the smoothness of the potential energy surface in the context
of \emph{ab initio} molecular dynamics (AIMD) simulations with hybrid
functionals.
\end{abstract}

\section{Introduction} \label{sec:intro}

Orbital pairs of the form
$\{\varphi_{i}(\vr)\psi_{j}(\vr)\}_{i,j=1}^{N}$, where
$\varphi_{i},\psi_{j}$ are single particle orbitals, appear
ubiquitously in quantum chemistry. A few examples include the Fock
exchange operator, the MP2 amplitude, and the polarizability
operator.\cite{SzaboOstlund1989,Martin2004} When $N$ is proportional
to the number of electrons $N_{e}$ in the system, the total number
of orbital pairs is $N^2\sim \Or(N_{e}^2)$. On the other hand, the
number of degrees of freedom needed to resolve all orbital pairs on
a dense grid is only $\Or(N_{e})$.  Hence as $N_{e}$ becomes large,
the set of all orbital pairs contains apparent redundant
information. In order to compress the redundant information and to
design more efficient numerical algorithms, many algorithms in the
past few decades have been developed. Pseudospectral
decomposition,\cite{MurphyJCP1995,ReynoldsJCP1996} Cholesky
decomposition,\cite{BeebeLinderberg1977,KochSanchezdePedersen2003,AquilantePedersenLindh2007,ManzerHornMardirossianEtAl2015}
density fitting (DF) or resolution of identity
(RI),\cite{NJP_14_053020_2012,Weigend2002} and tensor
hypercontraction (THC)\cite{ParrishHohensteinMartinezEtAl2012,
ParrishHohensteinMartinezEtAl2013} are only a few examples towards
this goal. When the single particle orbitals $\varphi_{i},\psi_{j}$
are already localized functions, ``local methods'' or ``linear
scaling methods''\cite{Goedecker1999, RPP_75_036503_2010_ON,
GuidonHutterVandevondele2010} can be applied to construct such
decomposition with cost that scales linearly with respect to
$N_{e}$. Otherwise, the storage cost of the matrix to represent all
orbital pairs on a grid is already $\Or(N_{e}^3)$, and the
computational cost of compressing the orbital pairs is then
typically $\Or(N_{e}^{4})$.

Recently, Lu and Ying developed a new decomposition called the
interpolative separable density fitting
(ISDF),\cite{JCP_302_329_2015_ISDF} which takes the following form
\begin{equation}
  \varphi_{i}(\vr)  \psi_{j}(\vr)\approx \sum_{\mu=1}^{N_{\mu}}
  \zeta_{\mu}(\vr) \left(\varphi_{i}(\hr{\mu})\psi_{j}(\hr{\mu})\right).
  \label{eqn:isdfformat}
\end{equation}
For a given $\vr$, if we view $\psi_{i}(\vr)\psi_{j}(\vr)$ as a row
of the matrix $\{\psi_{i}\psi_{j}\}$ discretized on a dense grid,
then the ISDF decomposition states that all such matrix rows can be
approximately expanded using a linear combination of matrix rows
with respect to a selected set of \textit{interpolation points}
$\{\hr{\mu}\}_{\mu=1}^{N_{\mu}}$.  The coefficients of such linear
combination, or \textit{interpolating vectors}, are denoted by
$\{\zeta_{\mu}(\vr)\}_{\mu=1}^{N_{\mu}}$. Here $N_{\mu}$ can be
interpreted as the numerical rank of the ISDF decomposition.
Compared to the standard density fitting method, the three-tensor
$\left(\varphi_{i}(\hr{\mu})\psi_{j}(\hr{\mu})\right)$ with three
indices $i,j,\mu$ takes a separable form. This reduces the storage
cost of the decomposed tensor from $\Or(N_{e}^3)$ to $\Or(N_{e}^2)$,
and the computational cost from $\Or(N_{e}^{4})$ to $\Or(N_{e}^3)$.
Note that if the interpolation points
$\{\hr{\mu}\}_{\mu=1}^{N_{\mu}}$ are chosen to be on a uniform grid,
then the ISDF decomposition reduces to the pseudospectral
decomposition, where $N_{\mu}\sim \Or(N_{e})$ but with a large
preconstant. For instance, the pseudospectral decomposition can be
highly inefficient for molecular systems, where the grid points in
the vacuum contributes nearly negligibly to the representation of
the orbital pairs. On the other hand, by selecting the interpolation
points carefully, e.g. through a randomized QR factorization with
column pivoting (QRCP) procedure~\cite{GolubVan2013}, the number of
interpolation points can be significantly reduced. The QRCP based
ISDF decomposition has been applied to accelerate a number of
applications such as two-electron integral
computation,\cite{JCP_302_329_2015_ISDF} correlation energy in the
random phase approximation,\cite{LuThicke2017} density functional
perturbation theory,\cite{LinXuYing2017} and hybrid density
functional calculations.\cite{JCTC_2017_ISDF} For example, in the
context of iterative solver for hybrid density functional
calculations, the Fock exchange operator $V_{X}$ defined in terms of
a set of orbitals $\{\varphi_i\}$ needs to be repeatedly applied to
another set of Kohn-Sham orbitals $\{\varphi_j\}$
\begin{equation}
  \left(V_{X}[\{\varphi_{i}\}]\psi_{j}\right)(\vr) =
  -\sum_{i=1}^{N_{e}} \varphi_{i}(\vr) \int K(\vr,\vr')
  \varphi_{i}(\vr')\psi_{j}(\vr') \ud \vr'.
  \label{eqn:applyVX}
\end{equation}
where $K(\vr,\vr')$ is the kernel for the Coulomb or the screened
Coulomb operator.  The integration in Eq.~\eqref{eqn:applyVX} is
often carried out by solving Poisson-like equations, using e.g. a
fast Fourier transform (FFT) method, and the computational cost is
$\Or(N_e^3)$ with a large preconstant. This is typically the most
time consuming component in hybrid functional calculations, and can
be accelerated by the ISDF decomposition for the orbital pairs
$\{\varphi_{i}\psi_{j}\}$.

In Ref.~\cite{JCP_302_329_2015_ISDF}, the interpolation points and
the interpolation vectors are determined simultaneously through a
randomized QR factorization with column pivoting (QRCP) applied to
$\{\psi_{i}(\vr)\psi_{j}(\vr)\}$ directly. We recently found that
the randomized QRCP procedure has $\Or(N_{e}^3)$ complexity but with
a relatively large preconstant, and may not be competitive enough
when used repeatedly. In order to overcome such difficulty, we
proposed a different approach in Ref.~\cite{JCTC_2017_ISDF} that
determines the two parts separately and reduces the computational
cost. We use the relatively expensive randomized QRCP procedure to
find the interpolation points in advance, and only recompute the
interpolation vectors whenever $\{\psi_{i}(\vr)\psi_{j}(\vr)\}$ has
been updated using an efficient least squares procedure that
exploits the separable nature of the matrix to be approximated.  As
a result, we can significantly accelerate hybrid functional
calculations using the ISDF decomposition in all but the first SCF
iteration.

In this work, we further remove the need of performing the QRCP
decomposition completely, and hence significantly reduce the
computational cost. Note that an effective choice of the set of
interpolation points should satisfy the following two conditions. 1) The
distribution of the interpolation points should roughly follow the
distribution of the electron density. In particular, there should be
more points when the electron density is high, and less or even zero
points if the electron density is very low. 2) The interpolation points
should not be very close to each other.  Otherwise matrix rows
represented by the interpolation points are nearly linearly dependent,
and the matrix formed by the interpolation vectors will be highly
ill-conditioned.  The QRCP procedure satisfies both 1) and 2)
simultaneously, and thus is an effective way for selecting the
interpolation points.  Here we demonstrate that 1) and 2) can also be
satisfied through a much simpler centroidal Voronoi tessellation (CVT)
procedure applied to a weight vector such as the electron density.

The Voronoi tessellation technique has been widely used in computer
science,\cite{aurenhammer1991voronoi} and scientific and engineering
applications such as image processing,\cite{du2006centroidal}
pattern recognition,\cite{ogniewicz1995hierarchic} and numerical
integration.\cite{becke1988multicenter} The concept of Voronoi
tessellation can be simply understood as follows. Given a discrete
set of weighted points, the CVT procedure divides a domain  into a
number of regions, each consisting of a collection of points that
are closest to its weighted centroid. Here we choose the electron
density as the weight, and the centroids as the interpolation
points. The centroids must be located where the electron density is
significant, and hence satisfy the requirement 1). The centroids are
also mutually separated from each other by a finite distance due to
the nearest neighbor principle, and hence satisfy the requirement
2). Although detailed analysis of the error stemming from such
choice of interpolation points is very difficult for general
nonlinear functions, we find that the CVT procedure approximately
minimizes the residual of the ISDF
decomposition~\eqref{eqn:isdfformat}. In practice, the CVT procedure
only applies to one vector (the electron density) instead of
$\Or(N_{e}^2)$ vectors and hence is very efficient.

We apply the ISDF-CVT method to accelerate hybrid functional
calculations in a planewave basis set. We perform such calculations
for different systems with insulating (liquid water), semiconducting
(bulk silicon), and metallic (disordered silicon aluminum alloy)
characters, as well as \textit{ab initio} molecular dynamics (AIMD)
simulations. We find that the ISDF-CVT method achieves similar
accuracy to that obtained from the ISDF-QRCP method,
with significantly improved efficiency. For instance, for a bulk
silicon system containing $1000$ silicon atoms computed on $2000$
computational cores, the QRCP procedure finds the interpolation
points with $434.2$ seconds, while the CVT procedure only takes
$3.2$ seconds. Since the solution of the CVT procedure is continuous
with respect to changes in the electron density, we also find that
the CVT procedure produces a smoother potential energy surface than
that by the QRCP procedure in the context of \emph{ab initio}
molecular dynamics (AIMD) simulations.


The remainder of the paper is organized as follows. We briefly
introduce the ISDF decomposition in section~\ref{sec:background}. In
section~\ref{sec:method} we describe the ISDF-CVT procedure and its
implementation for hybrid functional calculations. We present
numerical results of the ISDF-CVT method in
section~\ref{sec:numerical}, and conclude in
section~\ref{sec:conclusion}. We also provide the theoretical
justification of the CVT method in Appendix~\ref{sec:appendix}.

\section{Interpolative Separable Density Fitting (ISDF) decomposition}\label{sec:background}

In this section, we briefly introduce the ISDF
decomposition~\cite{JCP_302_329_2015_ISDF} evaluated using the method developed in
Ref.~\cite{JCTC_2017_ISDF}, which employs a separate treatment
of the interpolation points and interpolation vectors.

First, assume the interpolation points $\{\hr{\mu}\}_{\mu=1}^{N_{\mu}}$
are known, then the interpolation vectors can be efficiently evaluated
using a least squares method as follows. Using a linear algebra
notation, Eq.~\eqref{eqn:isdfformat} can be written as
\begin{equation}
  Z \approx \Theta C,
  \label{eqn:isdflineq}
\end{equation}
where each column of $Z$ is given by $Z_{ij}(\vr)=\varphi_{i}(\vr)\psi_{j}(\vr)$ sampled on a dense
real space grids $\{\vr_{i}\}_{i=1}^{N_g}$, and $\Theta = [\zeta_1,
\zeta_2, ..., \zeta_{N_{\mu}}]$ contains the interpolating vectors.
Each column of $C$ indexed by $(i,j)$ is given by
\[ [ \varphi_{i}(\hr{1})\psi_{j}(\hr{1}), \cdots,
    \varphi_{i}(\hr{\mu})\psi_{j}(\hr{\mu}), \cdots,
    \varphi_{i}(\hr{N_{\mu}})\psi_{j}(\hr{N_{\mu}})]^T.
\]
Eq.~\eqref{eqn:isdflineq} is an over-determined linear system with
respect to the interpolation vectors $\Theta$. The least squares
approximation to the solution is given by
\begin{equation}
\Theta = ZC^T (CC^T)^{-1}. \label{eq:Theta}
\end{equation}
It may appear that the matrix-matrix multiplications $ZC^T$ and
$CC^T$ take $\Or(N_{e}^{4})$ operations because the size of $Z$ is
$N_g \times (N_{e}N)$ and the size of $C$ is $N_{\mu} \times
(N_{e}N)$.  However, both multiplications can be carried out with
fewer operations due to the separable structure of $Z$ and $C$. The
computational complexity for computing the interpolation vectors is
$\Or(N_{e}^{3})$, and numerical results indicate that the
preconstant is also much smaller than that involved in hybrid
functional calculations.\cite{JCTC_2017_ISDF} Hence the
interpolation vectors can be obtained efficiently using the least
squares procedure.

The problem for finding a suitable set of interpolation points
$\{\hr{\mu}\}_{\mu=1}^{N_{\mu}}$ can be formulated as the following
linear algebra problem. Consider the discretized matrix $Z$ of size
$N_{g}\times N^2$, and find $N_{\mu}$ rows of $Z$
so that the rest of the rows of $Z$ can be approximated by the
linear combination of the selected $N_{\mu}$ rows. This is called an
interpolative decomposition\cite{SIAM_13_727_1992_QRCP}, and a
standard method to achieve such a decomposition is the QR
factorization with column pivoting (QRCP)
procedure\cite{SIAM_13_727_1992_QRCP} as
\begin{equation}
  Z^{T} \Pi = QR.
  \label{eqn:QRCP}
\end{equation}
Here $Z^T$ is the transpose of $Z$, $Q$ is an $N^2 \times N_g$
matrix that has orthonormal columns, $R$ is an upper triangular
matrix, and $\Pi$ is a permutation matrix chosen so that the
magnitude of the diagonal elements of $R$ form an non-increasing
sequence.  The magnitude of each diagonal element $R$ indicates how
important the corresponding column of the permuted $Z^T$ is, and
whether the corresponding grid point should be chosen as an
interpolation point. The QRCP factorization can be terminated when
the $(N_{\mu}+1)$-th diagonal element of $R$ becomes less than a
predetermined threshold. The leading $N_{\mu}$ columns of the
permuted $Z^T$ are considered to be linearly independent
numerically. The corresponding grid points are chosen as the
interpolation points. The indices for the chosen interpolation
points $\{\hr{\mu}\}$ can be obtained from indices of the nonzero
entries of the first $N_{\mu}$ columns of the permutation matrix
$\Pi$.

The QRCP decomposition satisfies the requirements 1) and 2) discussed in
the introduction. First, QRCP permutes matrix columns of $Z^{T}$ with large
norms to the front, and pushes matrix columns of $Z^{T}$ with small
norms to the back. Note that the square of the vector 2-norm of the column of
$Z^{T}$ labeled by $\vr$ is just
\begin{equation}
  \sum_{i,j=1}^{N} \varphi_{i}^2(\vr) \varphi_{j}^2(\vr) =
  \left(\sum_{i=1}^{N} \varphi_{i}^2(\vr)\right) \left(\sum_{j=1}^{N}
  \psi_{j}^2(\vr)\right).
  \label{}
\end{equation}
In the case when $\varphi_{i},\psi_{j}$ are the set of occupied
orbitals, the norm of each column of $Z^{T}$ is simply the electron
density. Hence the interpolation points chosen by QRCP will occur
where the electron density is significant. Second, once a column is
selected, all other columns are immediately orthogonalized with
respect to the chosen column. Hence nearly linearly dependent matrix
columns will not be selected repeatedly. As a result, the
interpolation points chosen by QRCP are well separated spatially.

It turns out that the direct application of the QRCP
procedure~\eqref{eqn:QRCP} still requires $\Or(N_{e}^{4})$
computational complexity.  The key idea used in
Ref.~\cite{JCP_302_329_2015_ISDF} to lower the cost is to randomly
subsample columns of the matrix $Z$ to form a smaller matrix
$\wt{Z}$ of size $N_{g}\times \wt{N}_{\mu}$, where $\wt{N}_{\mu}$ is
only slightly larger than $N_{\mu}$.  Applying the QRCP procedure to
this subsampled matrix $\wt{Z}$ approximately yields the choice of
interpolation points, but the computational complexity is reduced to
$\Or(N_{e}^3)$. In the context of hybrid density functional
calculations, we demonstrated that the cost of the randomized QRCP
method can be comparable to that of applying the exchange operator
in the planewave basis set.\cite{JCTC_2017_ISDF} However, the ISDF
decomposition can still significantly reduce the computational cost,
since the interpolation points only need to be performed once for a
fixed geometric configuration.

\section{Centroidal Voronoi Tessellation based ISDF decomposition}\label{sec:method}


In this section, we demonstrate that the interpolation points can
also be selected from a Voronoi tessellation procedure. For a
$d$-dimensional space, the Voronoi tessellation partitions a set of
points $\{\vr_{i}\}_{i=1}^{N_{g}}$ in $\R^{d}$ into a number of
disjoint cells. The partition is based on the distance of each point
to a finite set of points, called its generators. In our context,
let $\{\hr{\mu}\}_{\mu=1}^{N_{\mu}}$ denote such a set of
generators, the corresponding cell of a given generator $\hr{\mu}$,
defined through a cluster of points $\mathcal{C}_\mu$ is
\begin{equation}
  \mathcal{C}_\mu = \{\vr_{i} ~\vert~\text{dist}(\vr_{i}\,,\, \hr{\mu}) <
  \text{dist}(\vr_{i}\, ,\, \hr{\nu}) \quad \text{for all}\quad \mu \neq
  \nu\}.
  \label{eqn:vcell}
\end{equation}
The distance can be chosen to be any metric, e.g. the $L^{2}$
distance as $\text{dist}(\vr\,,\, \vr')=\norm{\vr-\vr'}$. In the case
when the distances of a point $\vr$ to $\hr{\mu},\hr{\nu}$ are
exactly the same, we may arbitrarily assign $\vr$ to one of the
clusters.

The Centroidal Voronoi tessellation (CVT) is a specific type of
Voronoi tessellation in which the generator $\hr{\mu}$ is chosen to
be the centroid of its cell. Given a weight function $\rho(\vr)$
(such as the electron density), the centroid of a cluster
$\mathcal{C}_\mu$ is defined as
\begin{equation}
  \vc(\mathcal{C}_\mu) = \frac{\sum_{\vr_{j}\in \mathcal{C}_{\mu}}
  \vr_j\;\rho(\vr_j)}{\sum_{\vr_j\in \mathcal{C}_\mu}\rho(\vr_j)}.
\end{equation}
Combined with the $L^2$ distance, CVT can be viewed as a
minimization problem over both all possible partition of the cells and
the centroids as~\cite{MacQueen1967}
\begin{equation}
  \label{eqn:CVT}
  \{\mathcal{C}_\mu^\ast,\vc_{\mu}^{*}\} =
  \underset{\{\mathcal{C}_\mu,\vc_{\mu}\}}{\text{arg}\min}\;\sum_{\mu=1}^{N_{\mu}}
  \sum_{\vr_k\in \mathcal{C}_\mu}\rho(\vr_k)\|\vr_i-\vc_{\mu}\|^2,
\end{equation}
and the interpolation points are then chosen to be the minimizers
$\hr{\mu}=\vc_{\mu}(\mathcal{C}_\mu^\ast)=\vc_{\mu}^{*}$. Following the
discussion in the introduction, the electron density as the weight
function~\eqref{eqn:CVT} enforces that the interpolation points should
locate at points where the electron density is significant and hence
satisfies the requirement 1). Since the cells $\mathcal{C}_\mu^\ast$ are
disjoint, the centroids $\vc_{\mu}^{*}$ are also separated by a finite
distance away from each other and hence satisfies the requirement 2). In
Appendix~\ref{sec:appendix} we provide another theoretical justification
in the sense that the CVT method approximately minimizes the residual
error of the ISDF decomposition.


Many algorithms have been developed to efficiently compute the
Voronoi tessellation \cite{medvedev1986algorithm}. One most widely
used method is the Llyod's algorithm~\cite{lloyd1982least}, which in discrete case is equivalent to
the K-Means algorithm~\cite{MacQueen1967}. The K-Means
algorithm is an iterative method that greedily minimizes the
objective by taking alternating steps between $\{\mathcal{C}_\mu\}$
and $\{\vc_\mu\}$. In this work, we adopt a weighted version of the
K-Means algorithm, which is demonstrated in Algorithm~\ref{wkmean}.
Note that the K-Means algorithm can be straightforwardly
parallelized. We distribute the grid points evenly at the beginning.
The classification step is the most time consuming step, and can be
locally computed for each group of grid points. After this step, the
weighted sum and total weight of all clusters can be reduced from
and broadcast to all processors for the next iteration.

\begin{algorithm}
\caption{Weighted K-Means Algorithm to Find Interpolation Points for Density Fitting}\label{wkmean}
\SetKwInOut{Input}{Input}
\SetKwInOut{Output}{Output}
\SetKwRepeat{Do}{do}{while}
\SetKwIF{If}{ElseIf}{Else}{if}{}{else if}{else}{end if}

\Input{Grid points $\{\vr_i\}_{i=1}^{N_{g}}$, Weight function $\rho(\vr)$, Initial
centroids $\{\vc^{(0)}_\mu\}$}
\Output{Interpolation points $\{\hr{\mu}\}_{\mu=1}^{N_{\mu}}$}
\textbf{Set} $t\gets 0$

\Do{$\{\vc^{(t)}_\mu\}$ not converged and maximum steps not reached}{
  \textbf{Classification step:}
  \For{$i = 1$ \KwTo $N_g$}{
  Assign point $\vr_{i}$ to the cluster $\mathcal{C}^{(t)}_{\mu}$
  \textbf{if} $\vc^{(t)}_\mu$ is the closest centroid to $\vr_i$
  }
  \textbf{Update step:}
  \For{$\mu = 1$ \KwTo $N_\mu$}{
  $\vc^{(t+1)}_\mu \gets
  {\sum_{\vr_{j}\in \mathcal{C}^{(t)}_{\mu}}
  \vr_j\;\rho(\vr_j)}/{\sum_{\vr_j\in \mathcal{C}^{(t)}_\mu}\rho(\vr_j)}$
  }

  \textbf{Set} $t\gets t+1$
}
\For{$\mu = 1$ \KwTo $N_\mu$}{
  \textbf{Set} $\hr{\mu}\gets \vc^{(t)}_{\mu}$
}
\end{algorithm}

\begin{figure}[htbp]
\begin{center}
\includegraphics[width=0.8\textwidth]{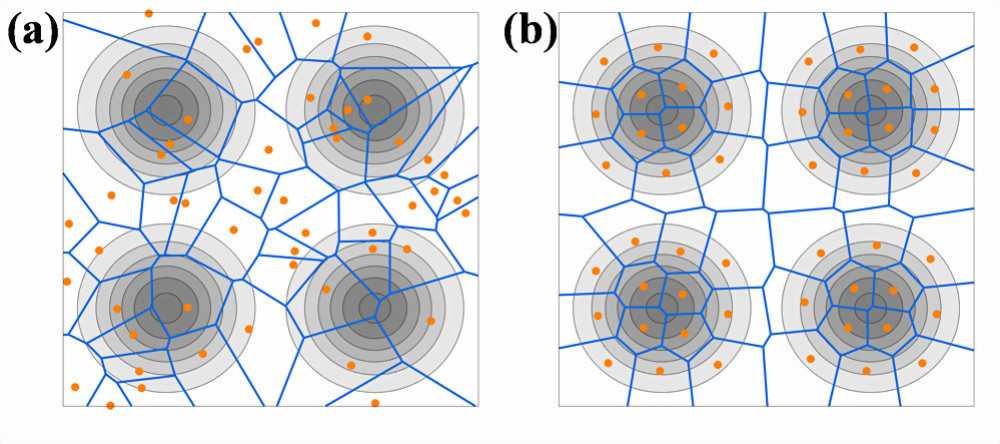}
\end{center}
\caption{Schematic illustration of the CVT procedure for 4 Gaussian
functions in a 2D domain, including (a) initial random choice of
centroids and Voronoi tessellation and centroidal Voronoi
tessellation generated by the weighted K-Means algorithm.}
\label{fig:CVT}
\end{figure}

In order to demonstrate the CVT procedure, we consider the weight
function $\rho(\vr)$ given by the summation of 4 Gaussian functions
in a 2D domain. The initial choice of centroids, given by 40
uniformly distributed random points, together with its
associated Voronoi tessellation are plotted in Figure~\ref{fig:CVT} (a).
Figure~\ref{fig:CVT} (b) demonstrates the converged centroids and the associated
Voronoi tessellation using the weighted K-Means algorithm. We observe that
the centroids concentrate on where the weight function is
significant, and are well-separated.

\begin{figure}[htbp]
\begin{center}
\includegraphics[width=0.45\textwidth]{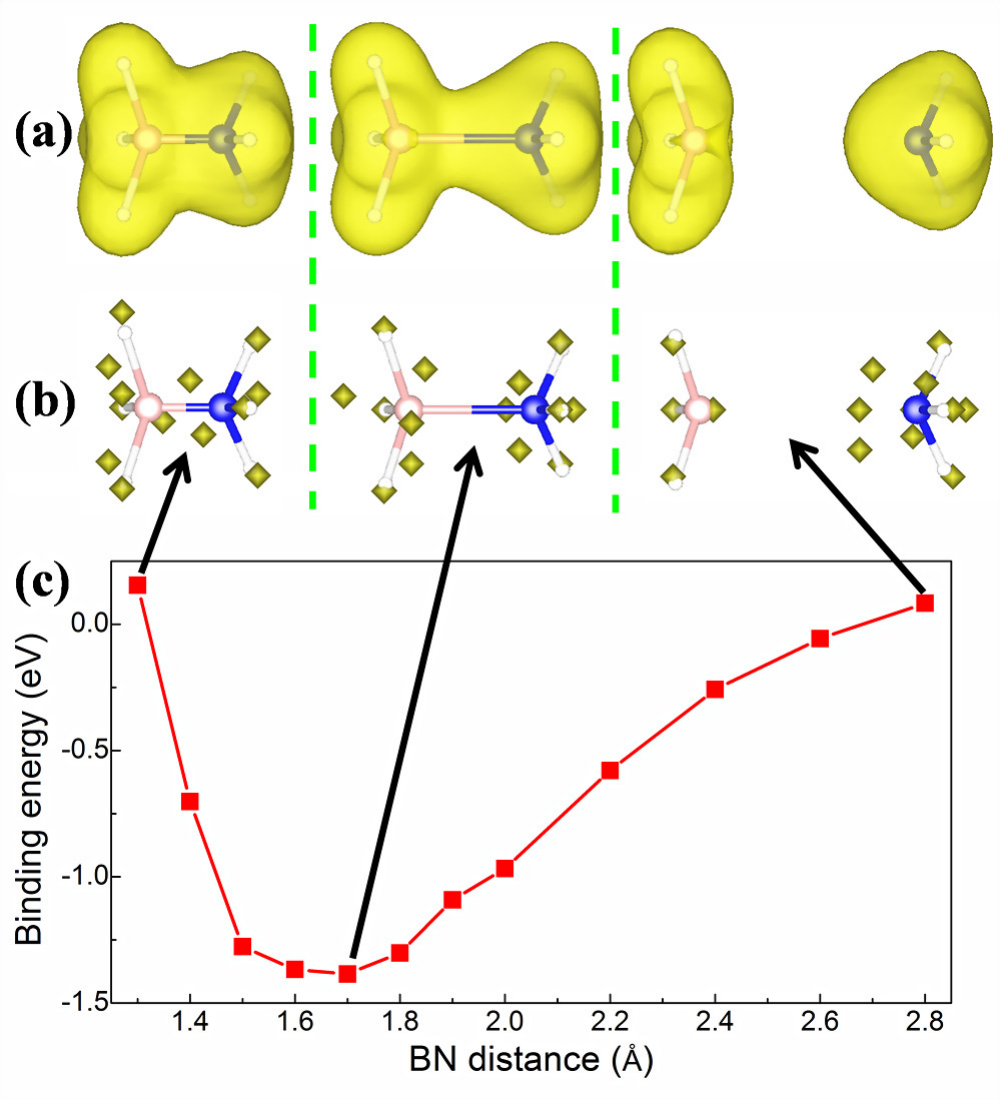}
\end{center}
\caption{The decomposition reaction process of BH$_3$NH$_3$ computed
with hybrid functional (HSE06) calculations by using the CVT
procedure to select interpolation points, including (a) the electron
density (yellow isosurfaces), (b) the interpolation points (yellow
squares) $\{\hr{\mu}\}_{\mu=1}^{N_{\mu}}$ ($N_{\mu}$ = 8) selected
from the real space grid points $\{\vr_{i}\}_{i=1}^{N_{g}}$ ($N_{g}$
= 66$^3$) when the BN distance respectively is 1.3, 1.7 and 2.8
{\AA} and (c) the binding energy as a function of BN distance for
BH$_3$NH$_3$ in a 10 {\AA} $\times$ 10 {\AA} $\times$ 10 {\AA} box.
The white, pink and blue pink balls denote hydrogen, boron and
nitrogen atoms, respectively.} \label{fig:BH3NH3}
\end{figure}

We also show how the interpolation points are placed and
moved during the ammonia-borane (BH$_3$NH$_3$) decomposition
reaction process. Figure~\ref{fig:BH3NH3} (a) shows the electron density
of the molecule at the compressed, equilibrium, and dissociated
configurations, respectively, according to the energy landscape in
Fig.~\ref{fig:BH3NH3} (c).
We plot the interpolation points found by the weighted K-Means algorithm
in Fig.~\ref{fig:BH3NH3} (b).
At the compressed configuration,
all the interpolation points are distributed evenly around the molecule.
As the bond length increases, some interpolation points are
transferred from BH$_3$ to NH$_3$. Finally at the dissociated
configuration, the NH$_3$ has more interpolation points around the
molecule, since there are more electrons in NH$_3$ than BH$_3$. Along
the decomposition reaction process, both the transfer of the
interpolation points and the potential energy landscape are smooth with
respect to the change of the bond length.

\section{Numerical results} \label{sec:numerical}

We demonstrate the accuracy and efficiency of the ISDF-CVT method
for hybrid functional calculations by using the DGDFT (Discontinuous
Galerkin Density Functional Theory) software
package.\cite{JCP_231_2140_2012_DGDFT, JCP_143_124110_2015_DGDFT,
PCCP_17_31397_2015_DGDFT, JCP_145_154101_2016_DGDFT,
JCP_335_426_2017_DGDFT} DGDFT is a massively parallel electronic
structure software package designed for large scale DFT calculations
involving up to tens of thousands of atoms. It includes a
self-contained module called PWDFT for performing planewave based
electronic structure calculations (mostly for benchmarking and
validation purposes). We implemented the ISDF-CVT method in PWDFT.
We use the Message Passing Interface (MPI) to handle data
communication, and the Hartwigsen-Goedecker-Hutter (HGH)
norm-conserving pseudopotential\cite{PRB_58_3641_1998_HGH}. All
calculations use the HSE06
functional.\cite{JCP_124_219906_2006_HSE06} All calculations are
carried out on the Edison systems at the National Energy Research
Scientific Computing Center (NERSC). Each node consists of two Intel
``Ivy Bridge'' processors with $24$ cores in total and 64 gigabyte
(GB) of memory. Our implementation only uses MPI. The number of
cores is equal to the number of MPI ranks used in the simulation.

In this section, we demonstrate the performance of the ISDF-CVT
method for accelerating hybrid functional calculations by using
three types of systems.\cite{JCTC_2017_PCDIIS} They consist of bulk silicon systems
(Si$_{64}$, Si$_{216}$ and Si$_{1000}$), a bulk water system with $64$ molecules ((H$_2$O)$_{64}$) and a
disordered silicon aluminum alloy system
(Al$_{176}$Si$_{24}$). Bulk silicon
systems (Si$_{64}$, Si$_{216}$ and Si$_{1000}$) and bulk water
system ((H$_2$O)$_{64}$) are semiconducting with a relatively large
energy gap $E_\text{gap}
> 1.0$ eV, and the Al$_{176}$Si$_{24}$ system is metallic with a
small energy gap $E_\text{gap} < 0.1$ eV. All
systems are closed shell systems, and the number of occupied bands
is $N_\text{band} = N_{e}/2$. In order to compute the energy gap in
the systems, we also include two unoccupied bands in all
calculations.

\subsection{Accuracy: Si$_{216}$ and Al$_{176}$Si$_{24}$}

We demonstrate the accuracy of the CVT-based ISDF decomposition in
hybrid functional calculation for semiconducting Si$_{216}$ and
metallic Al$_{176}$Si$_{24}$ systems, respectively. Although there
is no general theoretical guarantee for the convergence of the
K-Means algorithm and the convergence can depend sensitively on the
initialization\cite{arthur2006slow,arthur2007k}, we find that in the
current context,  initialization to have little impact on the final
accuracy of the approximation. Hence we use random initialization
for the K-Means algorithm. In all calculations, the adaptively
compressed exchange (ACE) technique is used to accelerate hybrid
functional calculations without loss of
accuracy\cite{JCTC_12_2242_2016_ACE}. The results obtained in this
work are labeled as ACE-ISDF (CVT), which are compared against those
obtained from the previous work based on the QRCP
decomposition~\cite{JCTC_2017_ISDF} labeled as ACE-ISDF (QRCP). In
both cases, we introduce a rank parameter $c$ to control the trade
off between efficiency and accuracy, by setting the number of
interpolation points $N_\mu = cN_e$. We measure the error using the
valence band maximum (VBM) energy level, the conduction band minimum
(CBM) energy level, the energy gap, the Hartree-Fock energy, the
total energy, and the atomic forces, respectively. The last three
quantities are defined as
\begin{align*}
\Delta{E_\text{HF}} &= |E_\text{HF}^\text{ACE-ISDF (CVT)} -
E_\text{HF}^\text{ACE}|/N_{A}\\
\Delta{E} &= |E^\text{ACE-ISDF (CVT)} - E^\text{ACE}|/N_{A}\\
{\Delta}F &= \max_I\|F_I^\text{ACE-ISDF (CVT)} - F_I^\text{ACE}\|
\end{align*}
where $N_A$ is the number of atoms and $I$ is the atom index.

\begin{figure}[htbp]
\begin{center}
\includegraphics[width=0.45\textwidth]{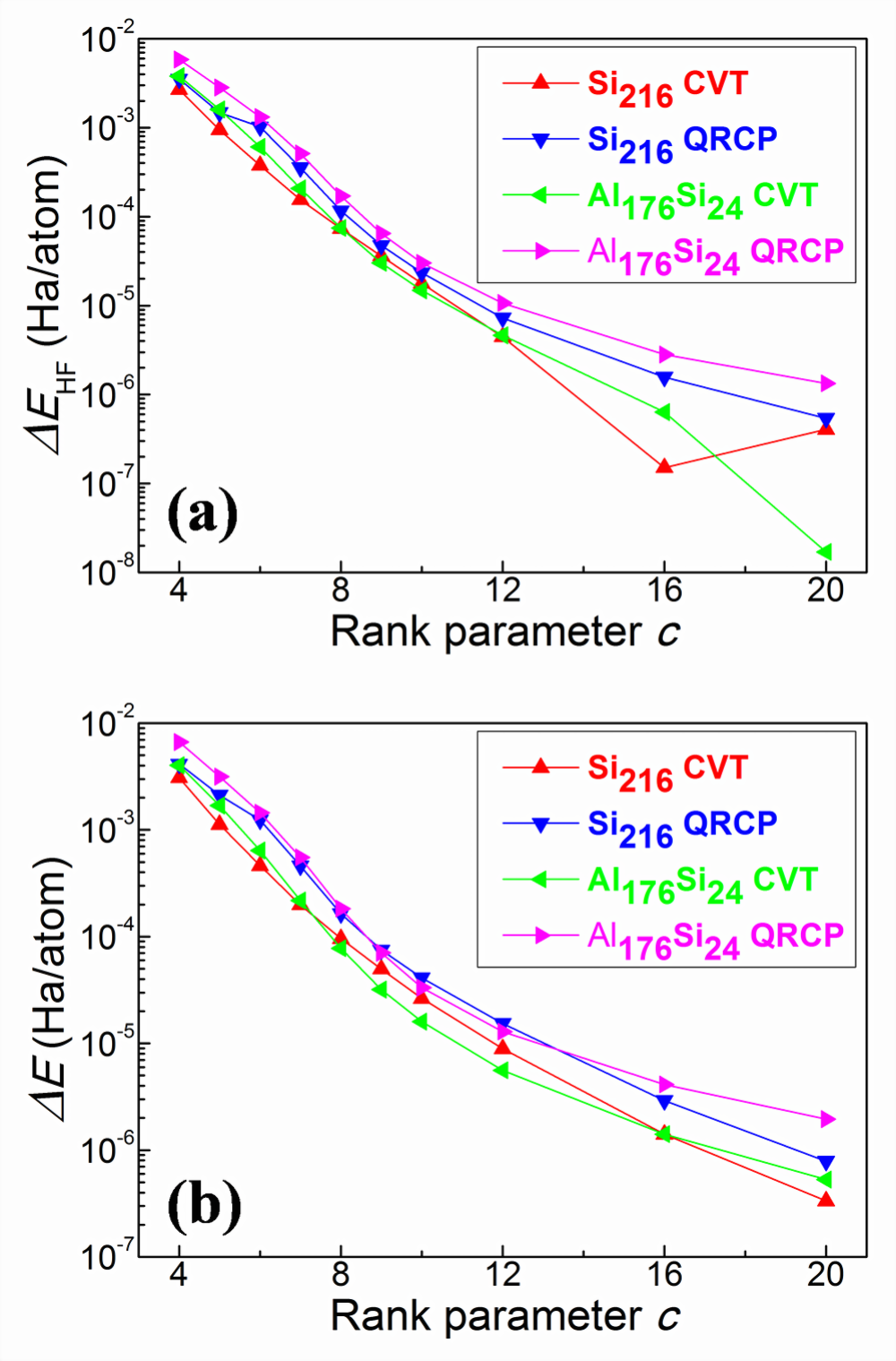}
\end{center}
\caption{The accuracy of ACE-ISDF based hybrid functional
calculations (HSE06) obtained by using the CVT and QRCP procedures
to select the interpolation points, with varying rank parameter $c$
from 4 to 20 for Si$_{216}$ and Al$_{176}$Si$_{24}$, including the
error of (a) Hartree-Fock energy ${\Delta}E_\text{HF}$ (Ha/atom) and
(b) total energy ${\Delta}E$ (Ha/atom).} \label{fig:Si216Al176Si24}
\end{figure}

Table~\ref{Accuracy} shows that the accuracy of the ACE-ISDF (CVT)
method can systematically improve as the rank parameter $c$
increases. When the rank parameter is large enough, the accuracy is 
fully comparable to that obtained from the benchmark
calculations. For a more modest choice $c=6.0$, the error of the
energy per atom reaches below the chemical accuracy of 1 kcal/mol
($1.6\times 10^{-3}$ Ha/atom), and the error of the force is around
$10^{-3}$ Ha/Bohr. This is comparable to the accuracy obtained from
ACE-ISDF (QRCP), and to e.g. linear scaling methods
for insulating systems with reasonable amount of truncation needed
to achieve significant speedup~\cite{JCTC_11_4655_2015}. In fact,
when compared with ACE-ISDF (QRCP) in
Figure~\ref{fig:Si216Al176Si24}, we find  that the CVT based ISDF
decomposition achieves slightly higher accuracy, though there is no
theoretical guarantee for this to hold in general. The last column
of Table~\ref{Accuracy} shows the runtime of the K-Means algorithm.
As $c$ increases, the number of interpolation points and hence the
number of cells increases proportionally. Hence we observe that the
runtime of K-Means scales linearly with respect to $c$.

\begin{table}
\caption{The accuracy of ACE-ISDF based hybrid functional
calculations (HSE06) obtained by using the CVT method to select
interpolation points, with varying rank parameter $c$ for
semiconducting Si$_{216}$ and metallic Al$_{176}$Si$_{24}$ systems.
The unit for VBM ($E_\text{VBM}$), CBM ($E_\text{CBM}$) and the
energy gap $E_\text{gap}$ is eV. The unit for the error in the
Hartree-Fock exchange energy ${\Delta}E_\text{HF}$ and the total
energy ${\Delta}E$ is Ha/atom, and the unit for the error in atomic
forces ${\Delta}F$ is Ha/Bohr. We use the results from the
ACE-enabled hybrid functional calculations as the reference. The last
column shows the time for K-Means with different $c$ values, with
434 cores for Si$_{216}$ and 314 cores for Al$_{176}$Si$_{24}$ on
Edison.} \label{Accuracy}
\begin{tabular}{cccccccc} \\ \hline \hline
\multicolumn{7}{c}{ACE-ISDF: Semiconducting Si$_{216}$ ($N_\text{band} = 432$)} \ \\
\hline
$c$  &  $E_\text{VBM}$ & $E_\text{CBM}$ & $E_\text{gap}$ & ${\Delta}E_\text{HF}$ & ${\Delta}E$ & ${\Delta}F$ & T\textsubscript{KMEANS}  \ \\
\hline
 4.0 & 6.7467 & 8.3433 & -1.5967 &  2.69E-03 &  3.08E-03 &  5.04E-03 &  0.228 \ \\
 5.0 & 6.6852 & 8.2231 & -1.5379 &  9.46E-04 &  1.12E-03 &  2.29E-03 &  0.248 \ \\
 6.0 & 6.6640 & 8.1522 & -1.4882 &  3.76E-04 &  4.62E-04 &  1.05E-03 &  0.301 \ \\
 7.0 & 6.6550 & 8.1163 & -1.4613 &  1.55E-04 &  1.98E-04 &  6.49E-04 &  0.312 \ \\
 8.0 & 6.6510 & 8.1030 & -1.4520 &  7.33E-05 &  9.55E-05 &  3.07E-04 &  0.349 \ \\
 9.0 & 6.6490 & 8.0980 & -1.4490 &  3.60E-05 &  4.96E-05 &  2.30E-04 &  0.398 \ \\
10.0 & 6.6479 & 8.0959 & -1.4480 &  1.78E-05 &  2.64E-05 &  1.30E-04 &  0.477 \ \\
12.0 & 6.6472 & 8.0945 & -1.4473 &  4.46E-06 &  8.91E-06 &  8.37E-05 &  0.530 \ \\
16.0 & 6.6469 & 8.0937 & -1.4468 &  1.51E-07 &  1.41E-06 &  3.20E-05 &  0.773 \ \\
20.0 & 6.6468 & 8.0935 & -1.4467 &  4.06E-07 &  3.33E-07 &  1.20E-05 &  0.830 \ \\
24.0 & 6.6468 & 8.0935 & -1.4467 &  2.99E-07 &  1.06E-07 &  5.18E-06 &  0.931 \ \\
 ACE & 6.6468 & 8.0934 & -1.4466 &  0.00E+00 &  0.00E+00 &  0.00E+00 &  0.000 \ \\
\hline \hline
\multicolumn{7}{c}{ACE-ISDF: Metallic Al$_{176}$Si$_{24}$ ($N_\text{band} = 312$)} \ \\
\hline
$c$  &  $E_\text{VBM}$ & $E_\text{CBM}$ & $E_\text{gap}$ & ${\Delta}E_\text{HF}$ & ${\Delta}E$ & ${\Delta}F$ & T\textsubscript{KMEANS} \ \\
\hline
 4.0 & 7.9258 & 8.0335 & -0.1076 &  3.80E-03 &  4.03E-03 &  8.01E-03 &  0.430 \ \\
 5.0 & 7.8537 & 7.9596 & -0.1059 &  1.60E-03 &  1.69E-03 &  3.18E-03 &  0.535 \ \\
 6.0 & 7.8071 & 7.9127 & -0.1056 &  6.07E-04 &  6.39E-04 &  1.48E-03 &  0.611 \ \\
 7.0 & 7.7843 & 7.8860 & -0.1017 &  2.07E-04 &  2.17E-04 &  1.03E-03 &  0.731 \ \\
 8.0 & 7.7749 & 7.8749 & -0.1000 &  7.43E-05 &  7.77E-05 &  4.40E-04 &  0.948 \ \\
 9.0 & 7.7718 & 7.8710 & -0.0992 &  3.02E-05 &  3.20E-05 &  1.98E-04 &  0.947 \ \\
10.0 & 7.7709 & 7.8697 & -0.0989 &  1.48E-05 &  1.60E-05 &  1.80E-04 &  1.096 \ \\
12.0 & 7.7703 & 7.8690 & -0.0987 &  4.64E-06 &  5.60E-06 &  8.51E-05 &  1.305 \ \\
16.0 & 7.7702 & 7.8688 & -0.0986 &  6.35E-07 &  1.41E-06 &  3.24E-05 &  1.646 \ \\
20.0 & 7.7701 & 7.8687 & -0.0986 &  1.70E-08 &  5.30E-07 &  1.91E-05 &  2.037 \ \\
 ACE & 7.7701 & 7.8687 & -0.0986 &  0.00E+00 &  0.00E+00 &  0.00E+00 &  0.000 \ \\
\hline \hline
\end{tabular}
\end{table}

\subsection{Efficiency: Si$_{1000}$}\label{sec:si1000}

We report the efficiency of the ISDF-CVT method by performing hybrid
DFT calculations for a bulk silicon system with 1000 atoms
($N_\text{band}$ = 2000) on 2000 computational cores as shown in
Table~\ref{Efficiency}, with respect to various choices of the
kinetic energy cutoff ($E_{\text{cut}}$).   With the number of
interpolation points fixed at $N_\mu = 12000$, both QRCP and K-Means
scales linearly with the number of grid points $N_g$. Yet the
runtime of K-Means is around two orders of magnitude faster than
QRCP. The determination of interpolation vectors, which consists of
solving a least-square problem, previously costs a fifth of the ISDF
runtime but now becomes the dominating component in CVT-based ISDF
decomposition. Notice that the ISDF method allows us to reduce the
number of Poisson-like equations from $N_e^2 = 4\times 10^6$ to
$N_\mu = 12000$, which results in a significant speedup in terms of
the cost of the FFT operations.
\begin{table}
\caption{The wall clock time (in seconds) spent in the components of
the ACE-ISDF and ACE enabled hybrid DFT calculations related to the
exchange operator, for Si$_{1000}$ on 2002 Edison cores at different
$E_{\text{cut}}$ levels. Interpolation points are selected via
either the QRCP or CVT procedure with the same rank parameter $c$ =
6.0. $N_g$ is the number of grid points in real space.}
\label{Efficiency}
\begin{tabular}{ccccccccc} \\ \hline \hline
\multicolumn{2}{c}{Si$_{1000}$}      &  & & \multicolumn{3}{c}{ACE-ISDF}  & & ACE  \ \\
$E_{\text{cut}}$ &  $N_g$  & & &  IP\textsubscript{QRCP} & IP\textsubscript{KMEANS} &  IV (FFT)   & & FFT  \ \\
\hline
10  &  \ph74\textsuperscript{3}  & &   & \ph38.06 & 0.70 & \ph12.48 (0.33) & & \phantom{ }85.15 \ \\
20  &  104\textsuperscript{3} & &   & 126.39 & 1.24 & \ph{ }36.48 (0.71) & & 143.54 \ \\
30  &  128\textsuperscript{3} & &   & 240.87 & 2.03 & \ph{ }68.50 (1.43) & & 268.88 \ \\
40  &  148\textsuperscript{3} & &   & 434.16 & 3.26 & 108.18 (3.10) & & 783.27 \ \\
\hline \hline
\end{tabular}
\end{table}

\FloatBarrier

\subsection{AIMD: Si$_{64}$ and (H$_2$O)$_{64}$} \label{sec:conv}

In this section, we demonstrate the accuracy of the ACE-ISDF (CVT)
method in the context of AIMD simulations for a bulk silicon system
Si$_{64}$ under the NVE ensemble, and a liquid water system
(H$_2$O)$_{64}$ under the NVT ensemble, respectively. The MD time
step size is $1.0$ femtosecond (fs). For the Si$_{64}$ system,
the initial MD structure (initial temperature $T$ = 300 K) is
optimized by hybrid DFT calculations, and we perform the simulation
for 0.5 ps. For the
(H$_2$O)$_{64}$ system, we perform the simulation for 2.0 ps to
sample the radial distribution function after equilibrating the
system starting from a prepared initial
guess.\cite{JCP_141_084502_2014} We use a single level Nose-Hoover
thermostat\cite{JCP_81_511_1984_Nose, PRA_31_1695_1985_Hoover} at
$T$ = $295$ K, and the choice of mass of the Nose-Hoover thermostat
is $85000$ au.

\begin{figure}[htbp]
\begin{center}
\includegraphics[width=0.85\textwidth]{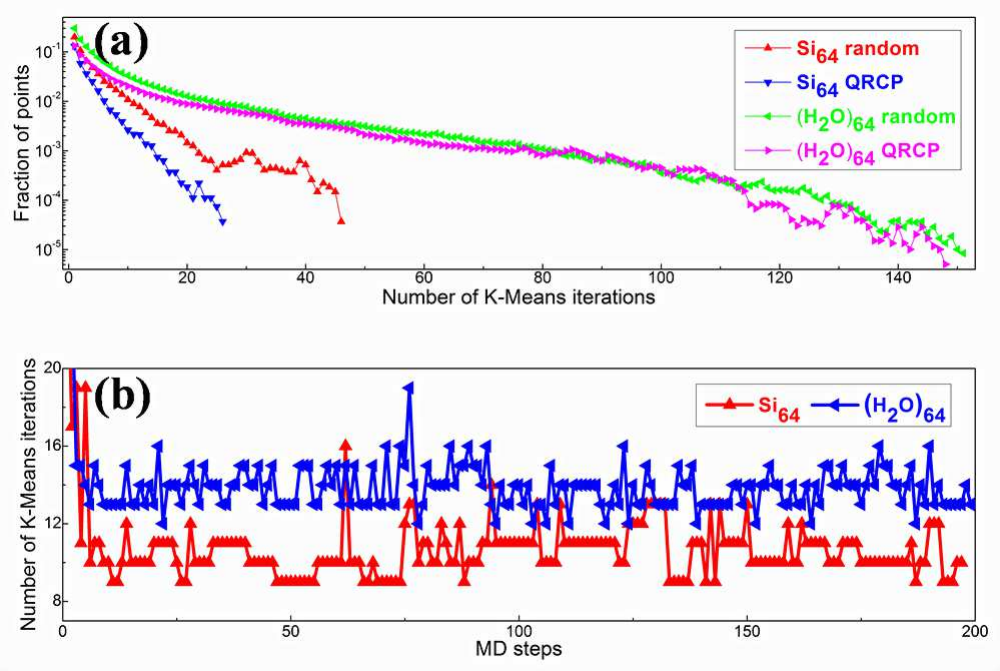}
\end{center}
\caption{Comparison of the ISDF-CVT method by using either random or
QRCP initialization for hybrid DFT AIMD simulations on bulk silicon
system Si$_{64}$ and liquid water system (H$_2$O)$_{64}$, including
(a) the fraction of points what switch cluster in each K-Means
iteration and (b) the number of K-Means iterations during each MD
step.} \label{fig:Si64H2O64MD}
\end{figure}

In the AIMD simulation, the interpolation points need to be
recomputed for each atomic configuration. At the initial MD step,
although the initialization strategy does not impact the accuracy of
the physical observable, it can impact the convergence
rate of the K-Means algorithm. We measure the convergence in terms
of the fraction of points that switch clusters during two
consecutive iterations. Figure~\ref{fig:Si64H2O64MD} (a) shows the
convergence of the K-Means algorithm with interpolation points
initially chosen from a random distribution and from the QRCP
solution, respectively. We find that the K-Means algorithm spends
around half the number of iterations to wait for $0.1\%$ of the
points to settle on the respective clusters. However, these points
often belong to the boundary of the clusters and have little effect
on the positions of the centroids (interpolation points). Therefore,
we decide to terminate K-Means algorithm whenever the fraction of
points that switch clusters falls below the $0.1\%$ threshold. It is
evident that QRCP initialization leads to faster convergence than
random sampling. However, in the AIMD simulation, a very good
initial guess of the interpolation points can be simply obtained
from those from the previous MD step. Figure~\ref{fig:Si64H2O64MD}
(b) shows that the number of K-Means iterations in the MD simulation
can be very small, which demonstrates the effectiveness of this
initialization strategy.

%

\begin{figure}[htbp]
\begin{center}
\includegraphics[width=0.85\textwidth]{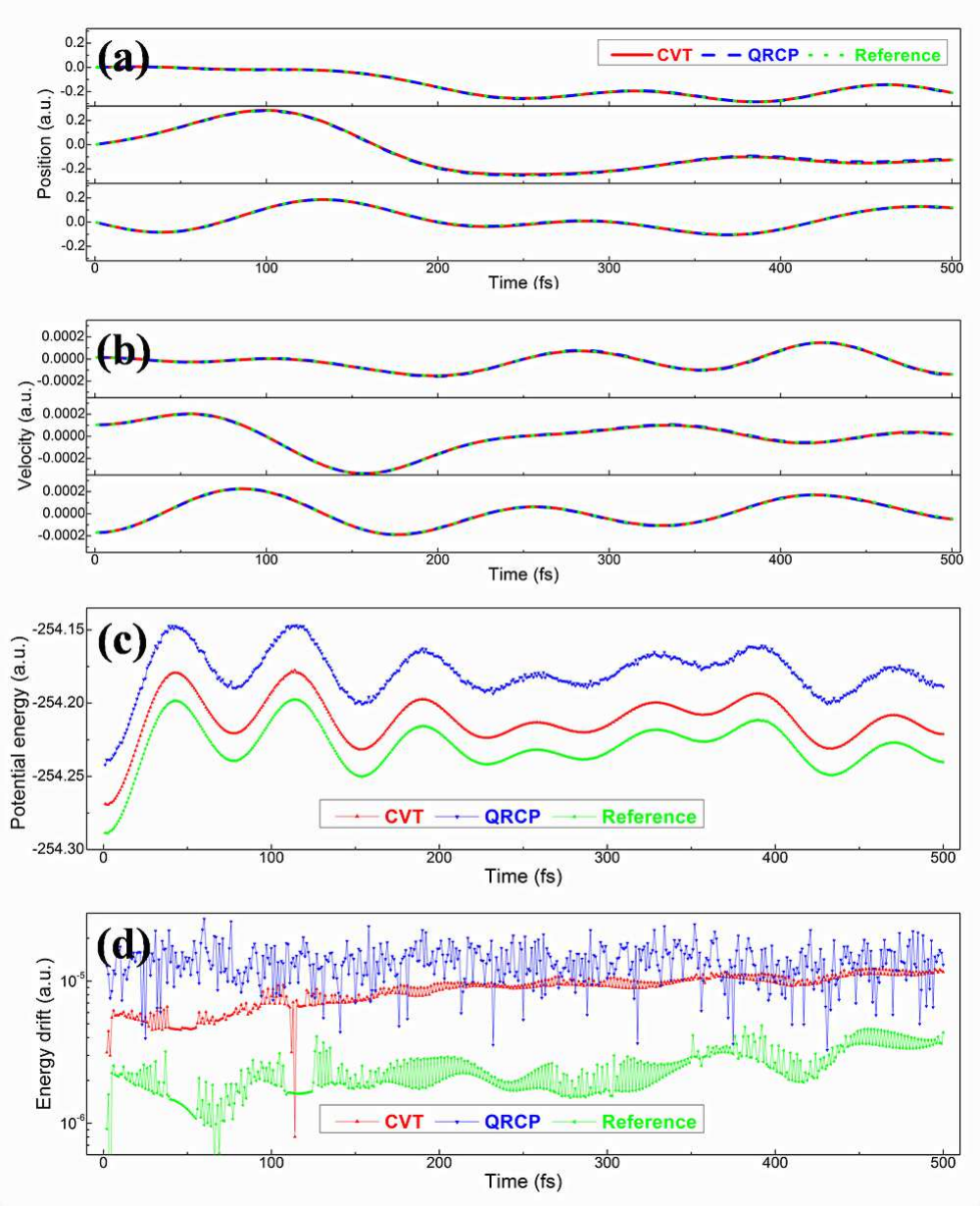}
\end{center}
\caption{Comparison of hybrid HSE06 DFT AIMD simulations by using
the ISDF-CVT and ISDF-QRCP methods as well as exact nested two-level
SCF iteration procedure as the reference on the bulk silicon
Si$_{64}$, including three coordinates (X, Y and Z directions) of
(a) position and (b) velocity of a specific Si atom, (c) potential
energy and (d) relatively energy drift during MD steps.}
\label{fig:Si64MD}
\end{figure}

Figure~\ref{fig:Si64MD} (a-b) demonstrate the positive and velocity
of a Si atom over a MD trajectory of $500$ fs obtained using the
ACE, ACE-ISDF (QRCP) and ACE-ISDF (CVT) methods (rank parameter $c$
= 8.0), respectively. The three trajectories fully overlap with each
other, indicating that ISDF is a promising method for reducing the
cost of hybrid functional calculations with controllable loss of
accuracy. Figure~\ref{fig:Si64MD} (c) shows the total potential
energy obtained by the three methods along the MD trajectory, and
the difference among the three methods is more noticeable.  This is
due to the fact that ISDF decomposition is a low rank decomposition
for the pair product of orbitals, which leads to error in the Fock
exchange energy and hence the total potential energy.  Nonetheless,
we find that such difference merely results in a shift of the
potential energy surface along the MD trajectory, and hence affects
little physical observables defined via relative potential energy
differences. Furthermore, the CVT method yields a potential energy
trajectory that is much smoother compared to that obtained from
QRCP. This is because the interpolation points obtained from CVT are
driven by the electron density, which varies smoothly along the MD
trajectory. Such properties do not hold for the QRCP method. This
means that the CVT method can be more effective when a smooth
potential energy surface is desirable, such as the case of geometry
optimization. The absolute error of the potential energy from the
CVT method is coincidentally smaller than that from QRCP, but we are
not aware of reasons for this behavior to hold in general.  Finally,
Figure~\ref{fig:Si64MD} (d) shows that both the CVT-based and
QRCP-based ISDF decomposition lead to controlled energy drift in the
NVE simulation.

\begin{figure}[htbp]
\begin{center}
\includegraphics[width=0.45\textwidth]{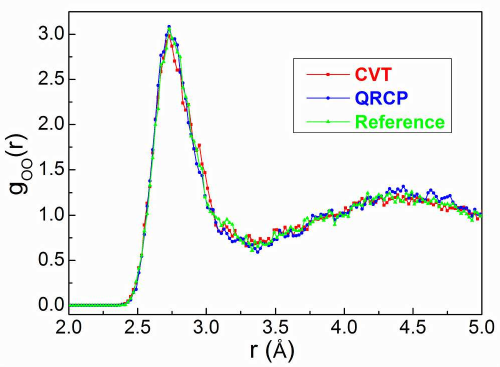}
\end{center}
\caption{The oxygen-oxygen radial distribution functions
$g_\text{OO}$($r$) of liquid water system (H$_2$O)$_{64}$ at $T$ =
$295$ K obtained from hybrid DFT AIMD simulations with the ISDF-CVT
and ISDF-QRCP methods as well as exact nested two-level SCF
iteration procedure as the reference.} \label{fig:H2O64gOO}
\end{figure}

We also apply the ACE-ISDF (CVT) and ACE-ISDF (QRCP) methods for
hybrid DFT AIMD simulations on liquid water system (H$_2$O)$_{64}$
under the NVT ensemble to sample the radial distribution function in
Figure~\ref{fig:H2O64gOO}. We find that the results from all three
methods agree very well, and our result is in quantitative agreement
with previous hybrid functional DFT
calculations,\cite{JCP_141_084502_2014} where the remaining
difference with respect to the experimental result can be to a large
extent attributed to the nuclei quantum
effects.\cite{MorroneCar2008}

\section{Conclusion} \label{sec:conclusion}

In this work, we demonstrate that the interpolative separable
density fitting decomposition (ISDF) can be efficiently performed
through a separated treatment of interpolation points and
interpolation vectors. We find that the centroidal Voronoi
tessellation method (CVT) provides an effective choice of
interpolation points using only the electron density as the input
information.  The resulting choice of interpolation points are by
design inhomogeneous in the real space, concentrated at regions
where the electron density is significant, and are well separated
from each other. These are all key ingredients for obtaining a low
rank decomposition that is accurate and a well conditioned set of
interpolation vectors. We demonstrate that the CVT-based ISDF
decomposition can be an effective strategy for reducing the cost
hybrid functional calculations for large systems. The CVT-based
method achieves similar accuracy when compared with that obtained
from QRCP, with significantly improved efficiency.  For a supercell
containing $1000$ silicon atoms on $2000$ computational cores, the
cost of QRCP-based method for finding the interpolation points costs
$434.2$ seconds, while the CVT procedure only takes $3.2$ seconds.
Since the solution of the CVT method depends continuously with
respect to the electron density, we also find that the CVT method
produces a smoother potential energy surface than that by the QRCP
method in the context of \emph{ab initio} molecular dynamics
simulation. Our analysis indicates that it might be possible to further improve
the quality of the interpolation points by taking into account the
gradient information in the weight vector.  We also expect that the
CVT-based strategy can also be useful in other contexts where the ISDF
decomposition is applicable, such as ground state calculations with
rung-5 exchange-correlation functionals, and excited state calculations.
These will be explored in the future work.

\section{Acknowledgments}

This work was partly supported by the National Science Foundation
under grant No. DMS-1652330, the DOE under grant No. DE-SC0017867,
the DOE CAMERA project (L. L.), and by the
DOE Scientific Discovery through Advanced Computing (SciDAC) program
(K. D., W.  H. and L. L.). The authors thank the National Energy Research
Scientific Computing (NERSC) center and the Berkeley Research
Computing (BRC) program at the University of California, Berkeley
for making computational resources available. We thank Anil Damle
and Robert Saye for useful discussions.

\appendix

\section{Minimization of the approximate residual} \label{sec:appendix}

The ISDF decomposition is a highly nonlinear process, and in general we
cannot expect the choice of interpolation points from CVT decomposition
to maximally reduce the error of the decomposition. Here
we demonstrate that the choice of the interpolation points from the
centroidal Voronoi tessellation algorithm approximately minimizes the
residual for the ISDF decomposition, and hence provides a heuristic
solution to the problem of finding interpolation points.

For simplicity we assume
$\varphi_{i}=\psi_{i}$, and hence each row of $Z$ is
$Z(\vr)=[\varphi_{i}(\vr)\varphi_{j}(\vr)]_{i,j=1}^N$.
Now suppose we cluster all matrix rows of $Z$ into
sub-collections $\{\C_\mu\}_{\mu=1}^{N_\mu}$, and for each $\C_\mu$ we
choose a representative matrix row
$Z(\vr_\mu)$. Then the error of the ISDF can be approximately
characterized as
\begin{equation}
R = \sum_{\mu = 1}^{N_\mu}\sum_{\vr_k\in \C_\mu}\left\|Z(\vr_k)-
\text{Proj}_{\text{span}\{Z(\vr_{\mu})\}}Z(\vr_k) \right\|^2,
\end{equation}
where the projection is defined according to the $L^{2}$ inner product
as
\begin{equation}
  \text{Proj}_{\text{span}\{Z(\vr_{\mu})\}}Z(\vr_k) =
  \frac{Z(\vr_{k})\cdot Z(\vr_\mu)}{Z(\vr_\mu)\cdot Z(\vr_\mu)}
  Z(\vr_{\mu}).
  \label{}
\end{equation}
Let $\Phi$ be the $N_{g} \times N$ matrix with each row $\Phi(\vr) =
[\varphi_i(\vr)]_{i=1}^N$, then the electron density $\rho(\vr)$ is
equal to $\Phi(\vr)\cdot \Phi(\vr)$.
Using the relation
\begin{equation}
  Z(\vr_\mu)\cdot Z(\vr_\mu) = (\Phi(\vr_{\mu})\cdot \Phi(\vr_{\mu}))^2
  = \rho(\vr_{\mu})^2,
  \label{}
\end{equation}
we have
\begin{equation}
  R =  \sum_{\mu=1}^{N_\mu}\sum_{\vr_k\in \C_\mu}
  \rho(\vr_k)^2\left(1-\frac{(\Phi(\vr_{k})\cdot\Phi(\vr_{\mu}))^4}{\rho(\vr_k)^2\rho(\vr_\mu)^2}\right)
  = \sum_{\mu=1}^{N_\mu}\sum_{\vr_k\in \C_\mu}\rho(\vr_k)^2 [1-\cos^4(\theta(\vr_k,\vr_\mu))].
  \label{}
\end{equation}
Here $\theta(\vr_{k},\vr_{\mu})$ is the angle between the vectors
$\Phi(\vr_{k})$ and $\Phi(\vr_{\mu})$. Use the fact that
\begin{equation}
  \rho(\vr_k) [1-\cos^4(\theta(\vr_k,\vr_\mu))] \le
  2 \Phi(\vr_{k})\cdot \Phi(\vr_{k})
  \sin^{2}(\theta(\vr_k,\vr_\mu)) \le
  2 \norm{\Phi(\vr_{k})-\Phi(\vr_{\mu})}^2,
  \label{}
\end{equation}
we have
\begin{equation}
  R \le 2 \sum_{\mu=1}^{N_\mu}\sum_{\vr_k\in \C_\mu}\rho(\vr_k)
  \norm{\Phi(\vr_{k})-\Phi(\vr_{\mu})}^2 \approx
  2 \sum_{\mu=1}^{N_\mu}\sum_{\vr_k\in \C_\mu}\rho(\vr_k)
  \norm{\nabla_{\vr}\Phi(\vr_{\mu})}^2 \norm{\vr_{k}-\vr_{\mu}}^2.
  \label{}
\end{equation}
If we further neglect the spatial inhomogeneity of the gradient $\Phi(\vr)$, we
arrive at the minimization criterion for the centroidal Voronoi
decomposition.

%


\begin{mcitethebibliography}{45}
\providecommand*\natexlab[1]{#1}
\providecommand*\mciteSetBstSublistMode[1]{}
\providecommand*\mciteSetBstMaxWidthForm[2]{}
\providecommand*\mciteBstWouldAddEndPuncttrue
  {\def\EndOfBibitem{\unskip.}}
\providecommand*\mciteBstWouldAddEndPunctfalse
  {\let\EndOfBibitem\relax}
\providecommand*\mciteSetBstMidEndSepPunct[3]{}
\providecommand*\mciteSetBstSublistLabelBeginEnd[3]{}
\providecommand*\EndOfBibitem{}
\mciteSetBstSublistMode{f}
\mciteSetBstMaxWidthForm{subitem}{(\alph{mcitesubitemcount})}
\mciteSetBstSublistLabelBeginEnd
  {\mcitemaxwidthsubitemform\space}
  {\relax}
  {\relax}

\bibitem[Szabo and Ostlund(1989)Szabo, and Ostlund]{SzaboOstlund1989}
Szabo,~A.; Ostlund,~N. \emph{{Modern Quantum Chemistry: Introduction to
  Advanced Electronic Structure Theory}}; McGraw-Hill, New York, 1989\relax
\mciteBstWouldAddEndPuncttrue
\mciteSetBstMidEndSepPunct{\mcitedefaultmidpunct}
{\mcitedefaultendpunct}{\mcitedefaultseppunct}\relax
\EndOfBibitem
\bibitem[Martin(2004)]{Martin2004}
Martin,~R. \emph{Electronic Structure -- Basic Theory and Practical Methods};
  Cambridge Univ. Pr.: West Nyack, {NY}, 2004\relax
\mciteBstWouldAddEndPuncttrue
\mciteSetBstMidEndSepPunct{\mcitedefaultmidpunct}
{\mcitedefaultendpunct}{\mcitedefaultseppunct}\relax
\EndOfBibitem
\bibitem[Murphy \latin{et~al.}(1995)Murphy, Beachy, Friesner, and
  Ringnalda]{MurphyJCP1995}
Murphy,~R.~B.; Beachy,~M.~D.; Friesner,~R.~A.; Ringnalda,~M.~N. Pseudospectral
  localized M{\o}ller–Plesset methods: Theory and calculation of
  conformational energies. \emph{J. Chem. Phys.} \textbf{1995}, \emph{103},
  1481\relax
\mciteBstWouldAddEndPuncttrue
\mciteSetBstMidEndSepPunct{\mcitedefaultmidpunct}
{\mcitedefaultendpunct}{\mcitedefaultseppunct}\relax
\EndOfBibitem
\bibitem[Reynolds \latin{et~al.}(1996)Reynolds, Martinez, and
  Carter]{ReynoldsJCP1996}
Reynolds,~G.; Martinez,~T.~J.; Carter,~E.~A. Local weak pairs spectral and
  pseudospectral singles and doubles configuration interaction. \emph{J. Chem.
  Phys.} \textbf{1996}, \emph{105}, 6455\relax
\mciteBstWouldAddEndPuncttrue
\mciteSetBstMidEndSepPunct{\mcitedefaultmidpunct}
{\mcitedefaultendpunct}{\mcitedefaultseppunct}\relax
\EndOfBibitem
\bibitem[Beebe and Linderberg(1977)Beebe, and Linderberg]{BeebeLinderberg1977}
Beebe,~N. H.~F.; Linderberg,~J. Simplifications in the generation and
  transformation of two-electron integrals in molecular calculations.
  \emph{Int. J. Quantum Chem.} \textbf{1977}, \emph{12}, 683\relax
\mciteBstWouldAddEndPuncttrue
\mciteSetBstMidEndSepPunct{\mcitedefaultmidpunct}
{\mcitedefaultendpunct}{\mcitedefaultseppunct}\relax
\EndOfBibitem
\bibitem[Koch \latin{et~al.}(2003)Koch, S{\'a}nchez~de Mer{\'a}s, and
  Pedersen]{KochSanchezdePedersen2003}
Koch,~H.; S{\'a}nchez~de Mer{\'a}s,~A.; Pedersen,~T.~B. Reduced scaling in
  electronic structure calculations using Cholesky decompositions. \emph{J.
  Chem. Phys.} \textbf{2003}, \emph{118}, 9481--9484\relax
\mciteBstWouldAddEndPuncttrue
\mciteSetBstMidEndSepPunct{\mcitedefaultmidpunct}
{\mcitedefaultendpunct}{\mcitedefaultseppunct}\relax
\EndOfBibitem
\bibitem[Aquilante \latin{et~al.}(2007)Aquilante, Pedersen, and
  Lindh]{AquilantePedersenLindh2007}
Aquilante,~F.; Pedersen,~T.~B.; Lindh,~R. {Low-cost evaluation of the exchange
  Fock matrix from Cholesky and density fitting representations of the electron
  repulsion integrals}. \emph{J. Chem. Phys.} \textbf{2007}, \emph{126},
  194106\relax
\mciteBstWouldAddEndPuncttrue
\mciteSetBstMidEndSepPunct{\mcitedefaultmidpunct}
{\mcitedefaultendpunct}{\mcitedefaultseppunct}\relax
\EndOfBibitem
\bibitem[Manzer \latin{et~al.}(2015)Manzer, Horn, Mardirossian, and
  Head-Gordon]{ManzerHornMardirossianEtAl2015}
Manzer,~S.; Horn,~P.~R.; Mardirossian,~N.; Head-Gordon,~M. {Fast, accurate
  evaluation of exact exchange: The occ-RI-K algorithm}. \emph{J. Chem. Phys.}
  \textbf{2015}, \emph{143}, 024113\relax
\mciteBstWouldAddEndPuncttrue
\mciteSetBstMidEndSepPunct{\mcitedefaultmidpunct}
{\mcitedefaultendpunct}{\mcitedefaultseppunct}\relax
\EndOfBibitem
\bibitem[Ren \latin{et~al.}(2012)Ren, Rinke, Blum, Wieferink, Tkatchenko,
  Sanfilippo, Reuter, and Scheffler]{NJP_14_053020_2012}
Ren,~X.; Rinke,~P.; Blum,~V.; Wieferink,~J.; Tkatchenko,~A.; Sanfilippo,~A.;
  Reuter,~K.; Scheffler,~M. Resolution-of-Identity Approach to {Hartree-Fock},
  Hybrid Density Functionals, {RPA}, {MP2} and {GW} with Numeric Atom-Centered
  Orbital Basis Functions. \emph{New J. Phys.} \textbf{2012}, \emph{14},
  053020\relax
\mciteBstWouldAddEndPuncttrue
\mciteSetBstMidEndSepPunct{\mcitedefaultmidpunct}
{\mcitedefaultendpunct}{\mcitedefaultseppunct}\relax
\EndOfBibitem
\bibitem[Weigend(2002)]{Weigend2002}
Weigend,~F. A fully direct {RI-HF} algorithm: Implementation, optimised
  auxiliary basis sets, demonstration of accuracy and efficiency. \emph{Phys.
  Chem. Chem. Phys.} \textbf{2002}, \emph{4}, 4285--4291\relax
\mciteBstWouldAddEndPuncttrue
\mciteSetBstMidEndSepPunct{\mcitedefaultmidpunct}
{\mcitedefaultendpunct}{\mcitedefaultseppunct}\relax
\EndOfBibitem
\bibitem[Parrish \latin{et~al.}(2012)Parrish, Hohenstein, Mart{\'\i}nez, and
  Sherrill]{ParrishHohensteinMartinezEtAl2012}
Parrish,~R.~M.; Hohenstein,~E.~G.; Mart{\'\i}nez,~T.~J.; Sherrill,~C.~D.
  {Tensor hypercontraction. II. Least-squares renormalization}. \emph{J. Chem.
  Phys.} \textbf{2012}, \emph{137}, 224106\relax
\mciteBstWouldAddEndPuncttrue
\mciteSetBstMidEndSepPunct{\mcitedefaultmidpunct}
{\mcitedefaultendpunct}{\mcitedefaultseppunct}\relax
\EndOfBibitem
\bibitem[Parrish \latin{et~al.}(2013)Parrish, Hohenstein, Mart{\'\i}nez, and
  Sherrill]{ParrishHohensteinMartinezEtAl2013}
Parrish,~R.~M.; Hohenstein,~E.~G.; Mart{\'\i}nez,~T.~J.; Sherrill,~C.~D.
  Discrete variable representation in electronic structure theory: Quadrature
  grids for least-squares tensor hypercontraction. \emph{J. Chem. Phys.}
  \textbf{2013}, \emph{138}, 194107\relax
\mciteBstWouldAddEndPuncttrue
\mciteSetBstMidEndSepPunct{\mcitedefaultmidpunct}
{\mcitedefaultendpunct}{\mcitedefaultseppunct}\relax
\EndOfBibitem
\bibitem[Goedecker(1999)]{Goedecker1999}
Goedecker,~S. {Linear scaling electronic structure methods}. \emph{Rev. Mod.
  Phys.} \textbf{1999}, \emph{71}, 1085--1123\relax
\mciteBstWouldAddEndPuncttrue
\mciteSetBstMidEndSepPunct{\mcitedefaultmidpunct}
{\mcitedefaultendpunct}{\mcitedefaultseppunct}\relax
\EndOfBibitem
\bibitem[Bowler and Miyazaki(2012)Bowler, and Miyazaki]{RPP_75_036503_2010_ON}
Bowler,~D.~R.; Miyazaki,~T. O(N) methods in Electronic Structure Calculations.
  \emph{Rep. Prog. Phys.} \textbf{2012}, \emph{75}, 036503\relax
\mciteBstWouldAddEndPuncttrue
\mciteSetBstMidEndSepPunct{\mcitedefaultmidpunct}
{\mcitedefaultendpunct}{\mcitedefaultseppunct}\relax
\EndOfBibitem
\bibitem[Guidon \latin{et~al.}(2010)Guidon, Hutter, and
  Vandevondele]{GuidonHutterVandevondele2010}
Guidon,~M.; Hutter,~J.; Vandevondele,~J. {Auxiliary density matrix methods for
  Hartree-Fock exchange calculations}. \emph{J. Chem. Theory Comput.}
  \textbf{2010}, \emph{6}, 2348--2364\relax
\mciteBstWouldAddEndPuncttrue
\mciteSetBstMidEndSepPunct{\mcitedefaultmidpunct}
{\mcitedefaultendpunct}{\mcitedefaultseppunct}\relax
\EndOfBibitem
\bibitem[Lu and Ying(2015)Lu, and Ying]{JCP_302_329_2015_ISDF}
Lu,~J.; Ying,~L. Compression of the Electron Repulsion Integral Tensor in
  Tensor Hypercontraction Format with Cubic Scaling Cost. \emph{J. Comput.
  Phys.} \textbf{2015}, \emph{302}, 329--335\relax
\mciteBstWouldAddEndPuncttrue
\mciteSetBstMidEndSepPunct{\mcitedefaultmidpunct}
{\mcitedefaultendpunct}{\mcitedefaultseppunct}\relax
\EndOfBibitem
\bibitem[Golub and Van~Loan(2013)Golub, and Van~Loan]{GolubVan2013}
Golub,~G.~H.; Van~Loan,~C.~F. \emph{Matrix computations}, 4th ed.; Johns
  Hopkins Univ. Press: Baltimore, 2013\relax
\mciteBstWouldAddEndPuncttrue
\mciteSetBstMidEndSepPunct{\mcitedefaultmidpunct}
{\mcitedefaultendpunct}{\mcitedefaultseppunct}\relax
\EndOfBibitem
\bibitem[Lu and Thicke(2017)Lu, and Thicke]{LuThicke2017}
Lu,~J.; Thicke,~K. {Cubic scaling algorithms for RPA correlation using
  interpolative separable density fitting}. \emph{J. Comput. Phys.}
  \textbf{2017}, \emph{351}, 187 -- 202\relax
\mciteBstWouldAddEndPuncttrue
\mciteSetBstMidEndSepPunct{\mcitedefaultmidpunct}
{\mcitedefaultendpunct}{\mcitedefaultseppunct}\relax
\EndOfBibitem
\bibitem[Lin \latin{et~al.}(2017)Lin, Xu, and Ying]{LinXuYing2017}
Lin,~L.; Xu,~Z.; Ying,~L. Adaptively Compressed Polarizability Operator for
  Accelerating Large Scale Ab Initio Phonon Calculations. \emph{Multiscale
  Model. Simul.} \textbf{2017}, \emph{15}, 29--55\relax
\mciteBstWouldAddEndPuncttrue
\mciteSetBstMidEndSepPunct{\mcitedefaultmidpunct}
{\mcitedefaultendpunct}{\mcitedefaultseppunct}\relax
\EndOfBibitem
\bibitem[Hu \latin{et~al.}(2017)Hu, Lin, and Yang]{JCTC_2017_ISDF}
Hu,~W.; Lin,~L.; Yang,~C. Interpolative Separable Density Fitting Decomposition
  for Accelerating Hybrid Density Functional Calculations With Applications to
  Defects in Silicon. \emph{J. Chem. Theory Comput.} \textbf{2017},
  \emph{accepted}\relax
\mciteBstWouldAddEndPuncttrue
\mciteSetBstMidEndSepPunct{\mcitedefaultmidpunct}
{\mcitedefaultendpunct}{\mcitedefaultseppunct}\relax
\EndOfBibitem
\bibitem[Aurenhammer(1991)]{aurenhammer1991voronoi}
Aurenhammer,~F. Voronoi diagrams—a survey of a fundamental geometric data
  structure. \emph{ACM Computing Surveys (CSUR)} \textbf{1991}, \emph{23},
  345--405\relax
\mciteBstWouldAddEndPuncttrue
\mciteSetBstMidEndSepPunct{\mcitedefaultmidpunct}
{\mcitedefaultendpunct}{\mcitedefaultseppunct}\relax
\EndOfBibitem
\bibitem[Du \latin{et~al.}(2006)Du, Gunzburger, Ju, and Wang]{du2006centroidal}
Du,~Q.; Gunzburger,~M.; Ju,~L.; Wang,~X. Centroidal Voronoi tessellation
  algorithms for image compression, segmentation, and multichannel restoration.
  \emph{Journal of Mathematical Imaging and Vision} \textbf{2006}, \emph{24},
  177--194\relax
\mciteBstWouldAddEndPuncttrue
\mciteSetBstMidEndSepPunct{\mcitedefaultmidpunct}
{\mcitedefaultendpunct}{\mcitedefaultseppunct}\relax
\EndOfBibitem
\bibitem[Ogniewicz and K{\"u}bler(1995)Ogniewicz, and
  K{\"u}bler]{ogniewicz1995hierarchic}
Ogniewicz,~R.~L.; K{\"u}bler,~O. Hierarchic voronoi skeletons. \emph{Pattern
  recognition} \textbf{1995}, \emph{28}, 343--359\relax
\mciteBstWouldAddEndPuncttrue
\mciteSetBstMidEndSepPunct{\mcitedefaultmidpunct}
{\mcitedefaultendpunct}{\mcitedefaultseppunct}\relax
\EndOfBibitem
\bibitem[Becke(1988)]{becke1988multicenter}
Becke,~A.~D. A multicenter numerical integration scheme for polyatomic
  molecules. \emph{The Journal of chemical physics} \textbf{1988}, \emph{88},
  2547--2553\relax
\mciteBstWouldAddEndPuncttrue
\mciteSetBstMidEndSepPunct{\mcitedefaultmidpunct}
{\mcitedefaultendpunct}{\mcitedefaultseppunct}\relax
\EndOfBibitem
\bibitem[Chan and Hansen(1992)Chan, and Hansen]{SIAM_13_727_1992_QRCP}
Chan,~T.~F.; Hansen,~P.~C. Some Applications of the Rank Revealing {QR}
  Factorization. \emph{SIAM J. Sci. Statist. Comput.} \textbf{1992}, \emph{13},
  727--741\relax
\mciteBstWouldAddEndPuncttrue
\mciteSetBstMidEndSepPunct{\mcitedefaultmidpunct}
{\mcitedefaultendpunct}{\mcitedefaultseppunct}\relax
\EndOfBibitem
\bibitem[MacQueen(1967)]{MacQueen1967}
MacQueen,~J. Some methods for classification and analysis of multivariate
  observations. Proc. of the Fifth Berkeley Symp. On Math. Stat. and Prob.
  1967; pp 281--297\relax
\mciteBstWouldAddEndPuncttrue
\mciteSetBstMidEndSepPunct{\mcitedefaultmidpunct}
{\mcitedefaultendpunct}{\mcitedefaultseppunct}\relax
\EndOfBibitem
\bibitem[Medvedev(1986)]{medvedev1986algorithm}
Medvedev,~N. The algorithm for three-dimensional Voronoi polyhedra.
  \emph{Journal of computational physics} \textbf{1986}, \emph{67},
  223--229\relax
\mciteBstWouldAddEndPuncttrue
\mciteSetBstMidEndSepPunct{\mcitedefaultmidpunct}
{\mcitedefaultendpunct}{\mcitedefaultseppunct}\relax
\EndOfBibitem
\bibitem[Lloyd(1982)]{lloyd1982least}
Lloyd,~S. Least squares quantization in PCM. \emph{IEEE transactions on
  information theory} \textbf{1982}, \emph{28}, 129--137\relax
\mciteBstWouldAddEndPuncttrue
\mciteSetBstMidEndSepPunct{\mcitedefaultmidpunct}
{\mcitedefaultendpunct}{\mcitedefaultseppunct}\relax
\EndOfBibitem
\bibitem[Lin \latin{et~al.}(2012)Lin, Lu, Ying, and E]{JCP_231_2140_2012_DGDFT}
Lin,~L.; Lu,~J.; Ying,~L.; E,~W. Adaptive Local Basis Set for Kohn¨CSham
  Density Functional Theory in a Discontinuous Galerkin Framework I: Total
  Energy Calculation. \emph{J. Comput. Phys.} \textbf{2012}, \emph{231},
  2140--2154\relax
\mciteBstWouldAddEndPuncttrue
\mciteSetBstMidEndSepPunct{\mcitedefaultmidpunct}
{\mcitedefaultendpunct}{\mcitedefaultseppunct}\relax
\EndOfBibitem
\bibitem[Hu \latin{et~al.}(2015)Hu, Lin, and Yang]{JCP_143_124110_2015_DGDFT}
Hu,~W.; Lin,~L.; Yang,~C. {DGDFT}: A Massively Parallel Method for Large Scale
  Density Functional Theory Calculations. \emph{J. Chem. Phys.} \textbf{2015},
  \emph{143}, 124110\relax
\mciteBstWouldAddEndPuncttrue
\mciteSetBstMidEndSepPunct{\mcitedefaultmidpunct}
{\mcitedefaultendpunct}{\mcitedefaultseppunct}\relax
\EndOfBibitem
\bibitem[Hu \latin{et~al.}(2015)Hu, Lin, and Yang]{PCCP_17_31397_2015_DGDFT}
Hu,~W.; Lin,~L.; Yang,~C. Edge Reconstruction in Armchair Phosphorene
  Nanoribbons Revealed by Discontinuous Galerkin Density Functional Theory.
  \emph{Phys. Chem. Chem. Phys.} \textbf{2015}, \emph{17}, 31397--31404\relax
\mciteBstWouldAddEndPuncttrue
\mciteSetBstMidEndSepPunct{\mcitedefaultmidpunct}
{\mcitedefaultendpunct}{\mcitedefaultseppunct}\relax
\EndOfBibitem
\bibitem[Banerjee \latin{et~al.}(2016)Banerjee, Lin, Hu, Yang, and
  Pask]{JCP_145_154101_2016_DGDFT}
Banerjee,~A.~S.; Lin,~L.; Hu,~W.; Yang,~C.; Pask,~J.~E. Chebyshev Polynomial
  Filtered Subspace Iteration in the Discontinuous Galerkin Method for
  Large-Scale Electronic Structure Calculations. \emph{J. Chem. Phys.}
  \textbf{2016}, \emph{145}, 154101\relax
\mciteBstWouldAddEndPuncttrue
\mciteSetBstMidEndSepPunct{\mcitedefaultmidpunct}
{\mcitedefaultendpunct}{\mcitedefaultseppunct}\relax
\EndOfBibitem
\bibitem[Zhang \latin{et~al.}(2017)Zhang, Lin, Hu, Yang, and
  Pask]{JCP_335_426_2017_DGDFT}
Zhang,~G.; Lin,~L.; Hu,~W.; Yang,~C.; Pask,~J.~E. Adaptive Local Basis Set for
  {Kohn–Sham} Density Functional Theory in a Discontinuous {Galerkin}
  Framework II: Force, Vibration, and Molecular Dynamics Calculations. \emph{J.
  Comput. Phys.} \textbf{2017}, \emph{335}, 426--443\relax
\mciteBstWouldAddEndPuncttrue
\mciteSetBstMidEndSepPunct{\mcitedefaultmidpunct}
{\mcitedefaultendpunct}{\mcitedefaultseppunct}\relax
\EndOfBibitem
\bibitem[Hartwigsen \latin{et~al.}(1998)Hartwigsen, Goedecker, and
  Hutter]{PRB_58_3641_1998_HGH}
Hartwigsen,~C.; Goedecker,~S.; Hutter,~J. Relativistic Separable Dual-Space
  Gaussian Pseudopotentials from {H to Rn}. \emph{Phys. Rev. B} \textbf{1998},
  \emph{58}, 3641\relax
\mciteBstWouldAddEndPuncttrue
\mciteSetBstMidEndSepPunct{\mcitedefaultmidpunct}
{\mcitedefaultendpunct}{\mcitedefaultseppunct}\relax
\EndOfBibitem
\bibitem[Heyd \latin{et~al.}(2006)Heyd, Scuseria, and
  Ernzerhof]{JCP_124_219906_2006_HSE06}
Heyd,~J.; Scuseria,~G.~E.; Ernzerhof,~M. Erratum: "Hybrid functionals based on
  a screened Coulomb potential" [{J. Chem. Phys.} 118, 8207 (2003)]. \emph{J.
  Chem. Phys.} \textbf{2006}, \emph{124}, 219906\relax
\mciteBstWouldAddEndPuncttrue
\mciteSetBstMidEndSepPunct{\mcitedefaultmidpunct}
{\mcitedefaultendpunct}{\mcitedefaultseppunct}\relax
\EndOfBibitem
\bibitem[Hu \latin{et~al.}(2017)Hu, Lin, and Yang]{JCTC_2017_PCDIIS}
Hu,~W.; Lin,~L.; Yang,~C. Projected Commutator DIIS Method for Accelerating
  Hybrid Functional Electronic Structure Calculations. \emph{J. Chem. Theory
  Comput.} \textbf{2017}, \emph{accepted}\relax
\mciteBstWouldAddEndPuncttrue
\mciteSetBstMidEndSepPunct{\mcitedefaultmidpunct}
{\mcitedefaultendpunct}{\mcitedefaultseppunct}\relax
\EndOfBibitem
\bibitem[Arthur and Vassilvitskii(2006)Arthur, and
  Vassilvitskii]{arthur2006slow}
Arthur,~D.; Vassilvitskii,~S. How slow is the k-means method? Proceedings of
  the twenty-second annual symposium on Computational geometry. 2006; pp
  144--153\relax
\mciteBstWouldAddEndPuncttrue
\mciteSetBstMidEndSepPunct{\mcitedefaultmidpunct}
{\mcitedefaultendpunct}{\mcitedefaultseppunct}\relax
\EndOfBibitem
\bibitem[Arthur and Vassilvitskii(2007)Arthur, and Vassilvitskii]{arthur2007k}
Arthur,~D.; Vassilvitskii,~S. k-means++: The advantages of careful seeding.
  Proceedings of the eighteenth annual ACM-SIAM symposium on Discrete
  algorithms. 2007; pp 1027--1035\relax
\mciteBstWouldAddEndPuncttrue
\mciteSetBstMidEndSepPunct{\mcitedefaultmidpunct}
{\mcitedefaultendpunct}{\mcitedefaultseppunct}\relax
\EndOfBibitem
\bibitem[Lin(2016)]{JCTC_12_2242_2016_ACE}
Lin,~L. Adaptively Compressed Exchange Operator. \emph{J. Chem. Theory Comput.}
  \textbf{2016}, \emph{12}, 2242--2249\relax
\mciteBstWouldAddEndPuncttrue
\mciteSetBstMidEndSepPunct{\mcitedefaultmidpunct}
{\mcitedefaultendpunct}{\mcitedefaultseppunct}\relax
\EndOfBibitem
\bibitem[Dawson and Gygi(2013)Dawson, and Gygi]{JCTC_11_4655_2015}
Dawson,~W.; Gygi,~F. Performance and Accuracy of Recursive Subspace Bisection
  for Hybrid DFT Calculations in Inhomogeneous Systems. \emph{J. Chem. Theory
  Comput.} \textbf{2013}, \emph{11}, 4655--4663\relax
\mciteBstWouldAddEndPuncttrue
\mciteSetBstMidEndSepPunct{\mcitedefaultmidpunct}
{\mcitedefaultendpunct}{\mcitedefaultseppunct}\relax
\EndOfBibitem
\bibitem[Jr. \latin{et~al.}(2014)Jr., Santra, Li, Wu, and
  Car]{JCP_141_084502_2014}
Jr.,~R. A.~D.; Santra,~B.; Li,~Z.; Wu,~X.; Car,~R. The Individual and
  Collective effects of Exact Exchange and Dispersion Interactions on the Ab
  Initio Structure of Liquid Water. \emph{J. Chem. Phys.} \textbf{2014},
  \emph{141}, 084502\relax
\mciteBstWouldAddEndPuncttrue
\mciteSetBstMidEndSepPunct{\mcitedefaultmidpunct}
{\mcitedefaultendpunct}{\mcitedefaultseppunct}\relax
\EndOfBibitem
\bibitem[Nos\'{e}(1984)]{JCP_81_511_1984_Nose}
Nos\'{e},~S. A Unified Formulation of the Constant Temperature Molecular
  Dynamics Methods. \emph{J. Chem. Phys.} \textbf{1984}, \emph{81}, 511\relax
\mciteBstWouldAddEndPuncttrue
\mciteSetBstMidEndSepPunct{\mcitedefaultmidpunct}
{\mcitedefaultendpunct}{\mcitedefaultseppunct}\relax
\EndOfBibitem
\bibitem[Hoover(1985)]{PRA_31_1695_1985_Hoover}
Hoover,~W.~G. Canonical Dynamics: Equilibrium Phase-Space Distributions.
  \emph{Phys. Rev. A} \textbf{1985}, \emph{31}, 1695\relax
\mciteBstWouldAddEndPuncttrue
\mciteSetBstMidEndSepPunct{\mcitedefaultmidpunct}
{\mcitedefaultendpunct}{\mcitedefaultseppunct}\relax
\EndOfBibitem
\bibitem[Morrone and Car(2008)Morrone, and Car]{MorroneCar2008}
Morrone,~J.; Car,~R. Nuclear quantum effects in water. \emph{Phys. Rev. Lett.}
  \textbf{2008}, \emph{101}, 017801\relax
\mciteBstWouldAddEndPuncttrue
\mciteSetBstMidEndSepPunct{\mcitedefaultmidpunct}
{\mcitedefaultendpunct}{\mcitedefaultseppunct}\relax
\EndOfBibitem
\end{mcitethebibliography}

\providecommand{\latin}[1]{#1}
\providecommand*\mcitethebibliography{\thebibliography}
\csname @ifundefined\endcsname{endmcitethebibliography}
  {\let\endmcitethebibliography\endthebibliography}{}

\end{document}